\documentclass[twocolumn]{aastex62}

\usepackage{bm}
\usepackage{soul}

\definecolor{RedOrange}{HTML}{F26035}

\definecolor{DarkGreen}{rgb}{0.13, 0.55, 0.13}

\graphicspath{{./}{figures/}}

\submitjournal{ApJ}

\shorttitle{Anisotropic Mass Segregation in Rotating Globular Clusters}
\shortauthors{Sz\"olgy\'en et al.}

\begin{document}

\title{\textbf{Anisotropic Mass Segregation in Rotating Globular Clusters}}

\correspondingauthor{\'Akos Sz\"olgy\'en}
\email{szolgyen@caesar.elte.hu}

\author[0000-0001-6062-2694]{\'Akos Sz\"olgy\'en}
\affil{Institute of Physics, E\"otv\"os University,
1/A P\'azm\'any P\'eter Street, Budapest, 1117, Hungary}

\author[0000-0003-3518-5183]{Yohai Meiron}
\affil{Institute of Physics, E\"otv\"os University,
1/A P\'azm\'any P\'eter Street, Budapest, 1117, Hungary}
\affiliation{Department of Astronomy and Astrophysics, University of Toronto,
50 Saint George Street, Toronto, ON M5S\,3H4, Canada}

\author[0000-0002-4865-7517]{Bence Kocsis}
\affil{Institute of Physics, E\"otv\"os University,
1/A P\'azm\'any P\'eter Street, Budapest, 1117, Hungary}

\begin{abstract}

We investigate the internal dynamics of anisotropic, rotating globular clusters with a multimass stellar population by performing new direct $N$-body simulations. In addition to the well-known radial mass segregation effect, where heavy stars and stellar remnants sink toward the center of the cluster, we find a mass segregation in the distribution of orbital inclinations as well. This newly discovered anisotropic mass segregation leads to the formation of a disk-like structure of massive objects near the equatorial plane of a rotating cluster. This result has important implications on the expected spatial distribution of black holes in globular clusters.
\end{abstract}


\section{Introduction} \label{sec:intro}

Galactic globular clusters (GCs) are dense, ancient stellar systems ($>$ 10 Gyr) that in many cases exhibit a significant amount of internal rotation \citep{Lane2011, Bellazzini2012, Bianchini2013, Fabricius2014, Kacharov2014, Kimmig2015, Lardo2015, Boberg2017, Jeffreson2017, Ferraro2018, Kamann2018, Lanzoni2018, Bianchini2018,Sollima2019}.
These star clusters represent a unique place for studying a variety of dynamical processes, such as two-body relaxation, mass segregation, stellar collisions, evaporation, and core collapse \citep{Meylan1997}.

Large number densities of stars facilitate close gravitational encounters in GCs. This defines them as collisional stellar systems in which pairwise encounters permit the exchange of orbital energies and angular momenta. This process, known as two-body relaxation, leads to diffusion of the phase space distribution function. As a consequence of the relaxation, the entropy increases and the system becomes inhomogeneous, forming a small, dense core of heavy objects and a large, low-density halo of light objects \citep{Binney2008}. This radial redistribution of stars, with respect to their masses, is the well-known radial mass segregation in stellar systems. The timescale of two-body relaxation ($t_{\mathrm{2b}}$) is roughly proportional to the number $N$ ($\approx 6 \times
10^{4} - 6 \times 10^{6}$), and the typical crossing time $t_{\mathrm{cross}}$ ($\approx 10^5 -10^6 $ yr) of stars in a Galactic GC \citep{Baumgardt2018,Binney2008}. The general estimate is $t_{\mathrm{2b}} \approx 0.1 N / \ln(N) \cdot t_{\mathrm{cross}} \approx 10^8 - 10^{10}$ yr. 
As $t_{\mathrm{2b}}$ is less than both the Hubble-time and the age of a typical GC, two-body relaxation plays a vital role in the dynamical evolution of GCs. 

Several theoretical studies showed that rotation may significantly affect the GCs' evolution. In their pioneering papers \citet{Einsel1999,Kim2002,Kim2004} investigated the dynamical evolution of rotating stellar systems by Fokker\---Planck models. In particular, \cite{Kim2004} studied rotating King models of GCs with mass spectra and found that both dynamical friction and initial rotation accelerate the dynamical evolution of GCs leading to a rapid core collapse. They showed that heavy objects segregate into the center as long as dynamical friction dominates in the competition with angular momentum exchange. The angular momentum of massive objects is sufficient to speed up their rotation causing gravogyro instability, a process during which angular momentum is transported outwards by the stellar dynamical analog of viscosity \citep{Hachisu1979,Hachisu1982}. 
They found that the heavy objects rotate faster than light ones in the center which leads to the suppression of mass segregation compared with the nonrotating GCs.

\citet{Ernst2007} performed direct $N$-body simulations of rotating GCs with both equal-mass and two-mass stellar populations. Their equal-mass models confirmed that rotation accelerates the dynamical evolution through the gravogyro instability, causing a faster contraction of the core due to the outward transport of angular momentum. They found that rotation acts in the opposite direction in the two-mass systems as it slows down radial mass segregation, which is in agreement with the findings of \cite{Kim2004}. Shortly after, \citet{Kim2007} also confirmed the acceleration of the core collapse in single-mass systems due to rotation. Later, \citet{Hong2013} published an extensive study on direct $N$-body simulations of GCs with different initial rotations with a two-mass model that showed qualitative agreement with the abovementioned Fokker\---Planck method results, and showed evidence for systematic angular momentum exchange between different mass components. \citet{Tiongco2016} have shown that the evolution in an external tidal field may naturally generate some rotation in GCs. Recently, \citet{Tiongco2017,Tiongco2018} broadened the above investigations by including velocity anisotropies with a range of external tidal field configurations and inclination angles with respect to the GC's internal angular momenta \citep[see also][]{Breen2017}. They found a variety of complex features in the evolution of the GCs' rotational properties including the possible formation of twisted differentially rotating geometries and counter-rotation between the inner and outer regions.

In a recent study, \cite{Meiron2018} investigated the relaxation and mixing of stars in $N$-body simulations of GCs, and found that persistent mutual gravitational torques can accelerate the relaxation of the orbital planes' distribution. 
This process, known as vector resonant relaxation (VRR), has been previously examined in the context of nuclear star clusters because there the competing two-body relaxation process is greatly suppressed due to the high velocity dispersion caused by the supermassive black hole \citep{Rauch1996,Hopman2006,Eilon2009,Kocsis2011,Kocsis2015,Fouvry2018b}. VRR clearly dominates the evolution in nuclear star clusters as it is two to four orders of magnitude faster than general two-body relaxation \citep{Kocsis2011,Kocsis2015}. As a consequence, the system can reach a thermodynamic equilibrium \citep{Roupas2017,Takacs2018,Fouvry2018a} that facilitates the formation of internal steady-state structures. Using statistical physics methods (i.e. Monte Carlo Markov Chains) \cite{Szolgyen2018} showed that VRR can lead to the formation of disk-like equilibrium structures of massive objects e.g.\ disks of heavy stars and stellar black holes in nuclear star clusters. This stochastic process is driven by resonant dynamical friction and causes anisotropic mass segregation \citep{Rauch1996}. During VRR, massive objects tend to relax to a ``disky'' configuration, while the distribution of low-mass objects becomes more spherical. The necessary condition for the formation of a massive stellar disk is the existence of an initially anisotropic multimass stellar population. Standard formation channels of galactic nuclei e.g. episodes of in-situ star formation and the infall of GCs can provide such anisotropy \citep{Szolgyen2018}. 

Motivated by these results, we explore if anisotropic mass segregation similarly operates in stellar clusters without a central massive object such as in GCs, particularly due to rotation.
We construct eight independent realizations of isolated GCs drawn from the distribution functions of rotating King models \citep{Longaretti1996} with different rotation parameters, number of particles, and mass spectra. We compare our results with a simulation without rotation and anisotropy to show that anisotropic mass segregation develops as a consequence of the clusters' internal rotation. In addition, we examine the evolution of an initially anisotropic cluster comprised of two counterrotating subsystems with which we demonstrate that anisotropic mass segregation may develop even in systems with zero-net rotation if it is initially anisotropic.

Using such toy models, we intentionally neglect the complexities of real systems (such as a realistic time-evolving mass function, stellar evolution, and binary evolution) in order to keep the uncertainties under control and to make a clean interpretation on the dynamical origin of anisotropic mass segregation. We follow the time evolution of the clusters using direct $N$-body simulations. We show that anisotropic mass segregation clearly appears in all initially anisotropic GCs.

\section{Methods} \label{sec:model}

We run a series of simulations using the phiGRAPE code \citep{Harfst2007}, which is a direct-summation $N$-body code that uses the Hermite integration scheme with block timesteps \citep{Makino1991}. We adopt rotating King models to generate initial conditions. This distribution function is defined as a functions of energy $E$ and the $z$-component of angular momentum $L_z$:
\begin{equation}
\label{eq:1}
    f(E,L_z) \propto \left( e^{-\beta E} - 1 \right) e^{-\beta \Omega_0 L_z} .
\end{equation}
There are two free parameters of the model: the angular velocity $\Omega_0$, and $\beta \equiv 1/(m\sigma^2)$, where $m$ is the average stellar mass, and $\sigma$ is the velocity dispersion of stars at the center of the cluster \citep{Einsel1999}. These can be transformed into dimensionless quantities, i.e., a rotation parameter $\omega_0 \equiv \Omega_0 \sqrt{9/(4 \pi G \rho_c)}$, where $\rho_c$ is the central density, $G$ is Newton's constant; and the King parameter $W_0 \equiv - \beta m( \Phi_0 - \Phi_t)$, where $\Phi_0$ is the central potential and $\Phi_t$ is the potential at the outer boundary of the model \citep{Ernst2007}. Our initial, rotating GCs are generated from such $f(E,L_z)$ rotating King distributions which are uniquely parameterized by $(W_0, \omega_0)$ pairs. All rotating clusters are initially flattened and the origin of the rotation is due to the anisotropic velocity distribution of stars. We adopt $N$-body units where the units of mass, length, and time are $M=\sum_i m_i$, $L=GM^2/(-4E)$, $T=GM^{5/2}/(-4E)^{3/2}$, where $E$ is the mechanical energy. In these units, $M$, $G$, and the virial radius ($R=M^{-2}\sum_{i\neq j}m_im_j/ |\bm{r}_i-\bm{r}_j| $) are unity for a system in virial equilibrium \citep{Henon1971,Heggie1986}. The clusters are initially in virial equilibrium but not in energy equipartition. We examine four different experiments with different initial conditions (see a summary of models in Table \ref{tab:1}):

(i) We perform five independent simulations with different rotation parameters: $\omega_0 = 0,0.3,0.6,0.7,0.8$ with the same $W_0 = 6$, following \cite{Einsel1999,Kim2002,Kim2004,Kim2007,Hong2013}. This choice for $W_0$ leads to a concentration parameter, $c=\log_{10} (r_t/r_0) = 1.2$ \citep{Binney2008}, which is consistent with several GCs in the \citet{Harris1996} catalog. We have not explored models with higher values of the rotation parameter because these models quickly evolve out of the dynamically stable regime showing a bar instability \citep{Hong2013}. In each simulation, the number of stars is set to be 64k ($k=1024$), the masses are drawn randomly from a mass distribution $p(m) \propto m^{-2}$ with a mass range defined by $m_{\mathrm{max}} /m_{\mathrm{min}} = 100$, and we use different random seeds for each realization. All five simulations are evolved up to 1000 time units, which proved to be sufficiently long to observe radial mass segregation due to two-body relaxation, and core collapse.

(ii) We investigate the dependence of the efficiency of anisotropic mass segregation on the number of stars. We generate four independent clusters with 32k, 64k, 128k, and 256k particles, with different random seeds but with the same initial rotation and King parameters $(W_0, \omega_0)=(6, 0.6)$, following \cite{Einsel1999,Kim2002,Kim2004,Kim2007,Hong2013} which has $v_\mathrm{rot}/\sigma \sim 0.86$ \citep{Einsel1999}. The mass distribution is the same as in (i) above. The simulations are evolved up to 2000 time units to guarantee reaching the two-body relaxation times for even the largest simulation (as it could be significantly longer for the larger $N$ runs). 

(iii) We also examine how the index of the power-law mass distribution affects the efficiency of anisotropic mass segregation. Fixing $(W_0, \omega_0)=(6, 0.6)$, as well as $N=64\mathrm{k}$, we compare the results of two models: one with $p(m) \propto m^{-1}$, and another with $p(m) \propto m^{-2}$ mass distribution in the same $m_{\mathrm{max}} /m_{\mathrm{min}} = 100$ ranges, such that $\sum_i m_i = 1$.

(iv) Finally, we investigate a superposition of two counterrotating King models. Both subclusters having 32k stars, $(W_0, \omega_0)=(6, 0.6)$ and $p(m) \propto m^{-2}$ mass distribution, but the total angular momenta were chosen to point in the opposite direction. Superimposing them yields a cluster of 64k stars with zero-net rotation, but with an axisymmetric initial structure. 

\begin{table}[ht]
\begin{center}
\begin{tabular}{lccccc}
\hline
model                                   & $N$           & $\omega_0$  & $v_\mathrm{rot}/\sigma$  & $\gamma$          & $W_0$ \\
\hline
M32.6.2     & 32k           & $0.6$     &  0.86  & $-2$         & $6$ \\
M64.0.2     & 64k           & $0.0$     &  0.00  & $-2$         & $6$ \\
M64.3.2     & 64k           & $0.3$     &  0.48  & $-2$         & $6$ \\
M64.6.1     & 64k           & $0.6$     &  0.86  & $-1$         & $6$ \\
M64.6.2     & 64k           & $0.6$     &  0.86  & $-2$         & $6$ \\
M64.7.2     & 64k           & $0.7$     &  0.97  & $-2$         & $6$ \\
M64.8.2     & 64k           & $0.8$     &  1.06  & $-2$         & $6$ \\
M128.6.2    & 128k          & $0.6$     &  0.86  & $-2$         & $6$ \\
M256.6.2    & 256k          & $0.6$     &  0.86  & $-2$         & $6$ \\
M64.6.2x    & $2\times 32$k & $0.6, -0.6$ & 0.00  & $-2$         & $6$ \\
\hline
\end{tabular}
\caption{\label{tab:1}Summary of $N$-body models examined in this paper. $N$ is the number of stars in a cluster, $\omega_0$ is the rotation parameter of the model, $v_\mathrm{rot}/\sigma$ is the ratio of the root-mean-squared rotational velocity to the velocity dispersion corresponding to the given $\omega_0$ \citep{Einsel1999,Ernst2007}, $\gamma$ $= \mathrm{d}\ln p/\mathrm{d}\ln m$, where $p$ is the mass distribution, and $W_0$ is the King parameter. Model M64.6.2x is composed of two counter-rotating M32.6.2. }
\end{center}
\vspace{-15pt}
\end{table}

In all experiments, the clusters were initialized with no binaries. The gravitational interactions were softened with a softening length of $3 \times 10^{-4}$ length units in order to prohibit the formation of binaries with a separation smaller than this length. This also helps to better conserve the energy of the system at the level of $\Delta E_\mathrm{tot} / E_\mathrm{tot} = 0.019$. The system conserves scalar angular momentum at the level of $\Delta L_\mathrm{tot} / L_\mathrm{tot} = 2 \times 10^{-5}$, and its direction by $10^{-6}$ rad.

\newpage

\section{Results} \label{sec:results}
\subsection{The Average Mass Enhancement}

To detect anisotropic mass segregation at any `snapshot' of the time evolution, we measure \textit{the enhancement of average mass}. This quantity is the average stellar mass in a bin of radius, $r$, and inclination cosine $\cos i = L_z/|\bm{L}|$, divided by the average stellar mass in the spherical shell at radius $r$. It can be written as 
\begin{equation}
\label{eq:2}
\varepsilon(r,\cos i)\equiv \frac{\bar{m}(r,\cos i)}{\bar{m}(r)},
\end{equation}
where $\bar{m}(r,\cos i)$ is the average mass of stars in a segment of a spherical shell around $r$, and $\cos i$; while $\bar{m}(r)$ is the average mass in a spherical shell around $r$. This essentially normalizes out the effect of radial mass segregation, revealing the relative effectiveness of anisotropic mass segregation for different values of $r$. 

To improve the statistics, we average the $\varepsilon(r,\cos i)$ of the last 300 snapshots out of 500 which means stacking snapshots together (corresponding to between 400 and 1000 time units). We note that this is reasonable, because as we show in Section \ref{sec:the_effect_of_rotation}, the system reaches equilibrium by this time. 

\begin{figure}[t!]
\includegraphics[width=1\columnwidth]{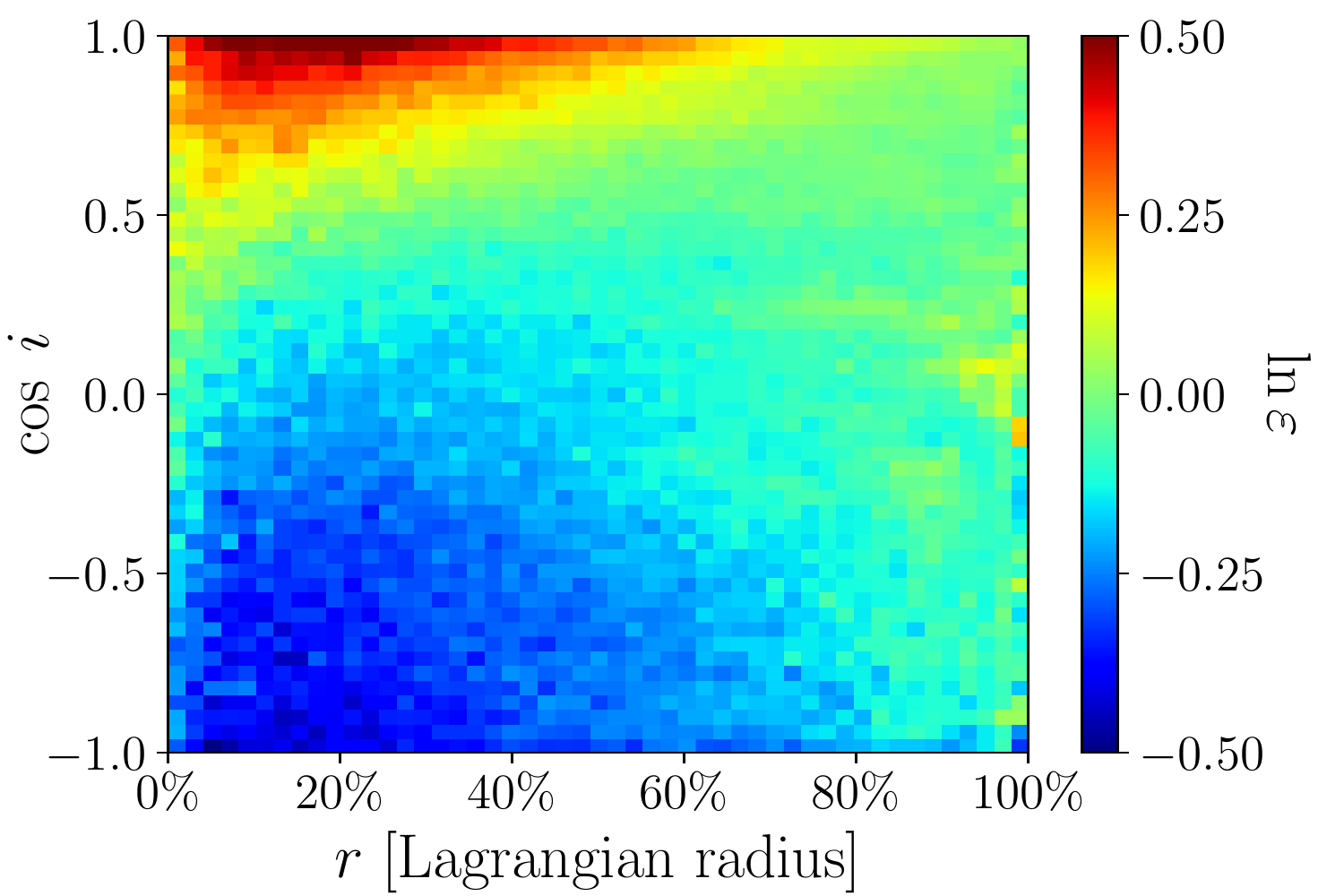}
\caption{\label{fig:1} Enhancement of average mass with respect to the average mass in spherical shells defined by Lagrangian radii in the  M64.6.2 model of a rotating globular cluster; see Equation (\ref{eq:2}) and Table \ref{tab:1}. The cluster is in an equilibrium state with respect to the distribution of inclinations. To improve statistics, this diagram is a stacking of snapshots between 400 and 1000 time units where the distribution of inclinations already reached an equilibrium.}
\end{figure}

The enhancement of average mass is shown in Figure~\ref{fig:1} for the $N=64\mathrm{k}$ model with $(W_0, \omega_0)=(6, 0.6)$ and an $m^{-2}$ mass function. The horizontal axis of Figure~\ref{fig:1} shows the distance from the cluster's center as the Lagrangian radius\footnote{Lagrangian radius expressed in percentage is corresponding to a physical radius of a sphere which encloses the given percentage of the total mass of the cluster}; this presentation has the advantage that the number of particles per radial bin is roughly constant by construction, and different models (as well as different snapshots in time for the same model) can be more easily compared. For this model, there is a significant enhancement of average mass at $\cos i\gtrsim 0.7$ which is most prominent at $\cos i\gtrsim 0.9$ around the $20\%$ Lagrangian radius, but which extends up to $\sim60\%$. This is a clear evidence of anisotropic mass segregation in $N$-body simulations of rotating GCs. 

\begin{figure*}[htp]
\includegraphics[width=2\columnwidth]{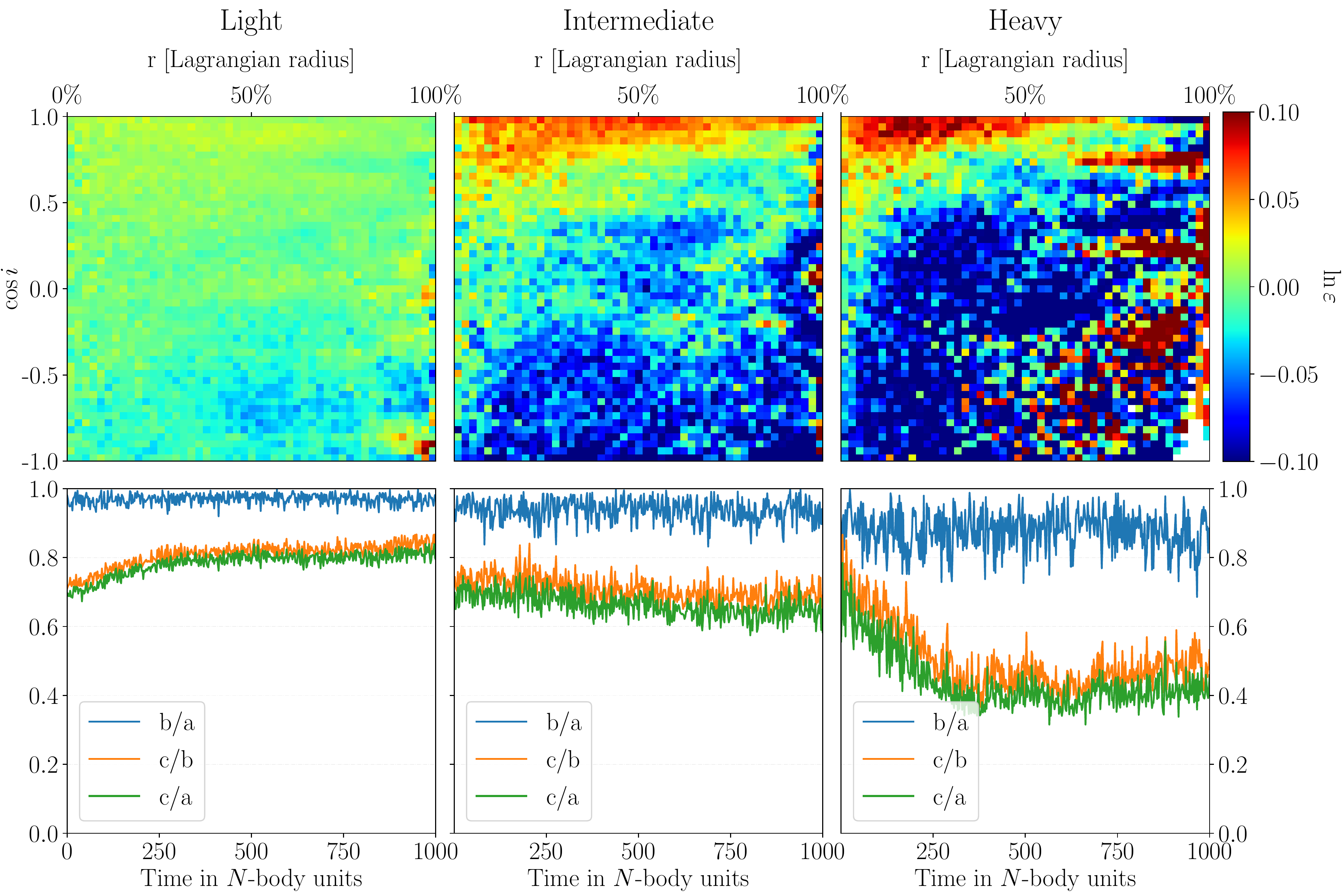}
\caption{\label{fig:2} Top panels are the same as in Figure~\ref{fig:1}, but the stellar population is divided into three mass groups: light, intermediate, and heavy, shown in panels from left to right. Note that a more refined color scale is used here. Anisotropic mass segregation is prominent in the intermediate and heavy groups. Bottom panels show the time evolution in the morphology of the cluster within the $50\%$ Lagrangian radius, particularly the ratios of the principal axes of the light, intermediate and heavy subpopulations. Here the axes $a\geq b \geq c$ are defined as the eigenvalues of the quadrupole moment tensor of the mass distribution of the given mass group.}
\end{figure*}

To examine the properties of anisotropic mass segregation further, we calculate the relative enhancement of average mass separately for three mass groups. To do so, we divide the stellar population into three mass groups uniformly on a logarithmic scale (i.e. $m_{\min}^{(4-i)/3}m_{\max}^{(i-1)/3} \leq m < m_{\min}^{(3-i)/3}m_{\max}^{i/3}$ for the three groups with $i=1,2,3$, respectively). We measure $\varepsilon(r,\cos i)$ for each group separately. With such separation $85.5\%$ of the stars are in the light group, $12.2\%$ in the intermediate, and $2.3\%$ in the heavy group, given the $m^{-2}$ mass spectrum. Figure~\ref{fig:2} shows that strong anisotropic mass segregation is present within the heavy population, it is weaker within the intermediate population, and nearly absent in the the light population.

The lower panels of Figure~\ref{fig:2} also show the axis ratios of these three mass groups as a function of time within the $50\%$ Lagrangian radius (where anisotropic mass segregation is the most prominent). We exclude escapers which are only $0.2\%$ of the whole cluster. Axes $a\geq b \geq c$ are defined as the eigenvalues of the quadrupole moment tensor of the mass distribution. The fact that $b/a \sim 1$ indicates that throughout the simulation the system remains axisymmetric and no significant triaxiality develops. The growth in $c/a$ for the light group indicates that the light subsystem becomes more spherical than initially, while the drop in $c/a$ for the heavy group indicates that heavy subsystem becomes more oblate than initially, which is consistent with the results drawn from inspection of the enhancement factor $\varepsilon$. This result is also in line with \citet{Kim2004}, who showed that the heavier component in their two-mass system evolved to rotate faster than the light component (see their Figure 8), indicating that it was probably more oblate as well. We also found that the dynamical evolution led to a $6.7\%$ increase in the total angular momentum within the half-mass radius which means that the rotation of the internal part sped up.

\newpage
\subsection{The Effect of Rotation} \label{sec:the_effect_of_rotation}

We can now characterize this diagram with a single number, in order to be able to compare the importance of anisotropic mass segregation in different models and different times for the same model. We choose to first marginalize over $r$ by averaging the $\varepsilon(r,\cos i)$ with the appropriate weights. We call the resulting quantity the \textit{effective enhancement},

\begin{equation}
\label{eq:3}
\tilde{\varepsilon}(\cos i)\equiv \int_0^{\infty} \varepsilon(r,\cos i) w(r) \; \mathrm{d}r,
\end{equation}
where the weight $w(r)$ is the relative number of particles in a spherical shell of thickness $\mathrm{d}r$ around $r$:
\begin{equation}
w(r)\equiv\frac{2\pi r^{2}}{N}\int_0^{\pi} n(r,\theta)\sin\theta \; \mathrm{d}\theta.
\end{equation}
Here, $n(r,\theta)$ is the number density (as a function of the radial distance $r$ and polar angle $\theta$) and $N$ is the total number of stars. The effective enhancement has the physical meaning of average mass of objects at inclination $i$. Objects are, on average, heavier on inclinations where this quantity is higher.

\begin{figure}[h!]
\includegraphics[width=1\columnwidth]{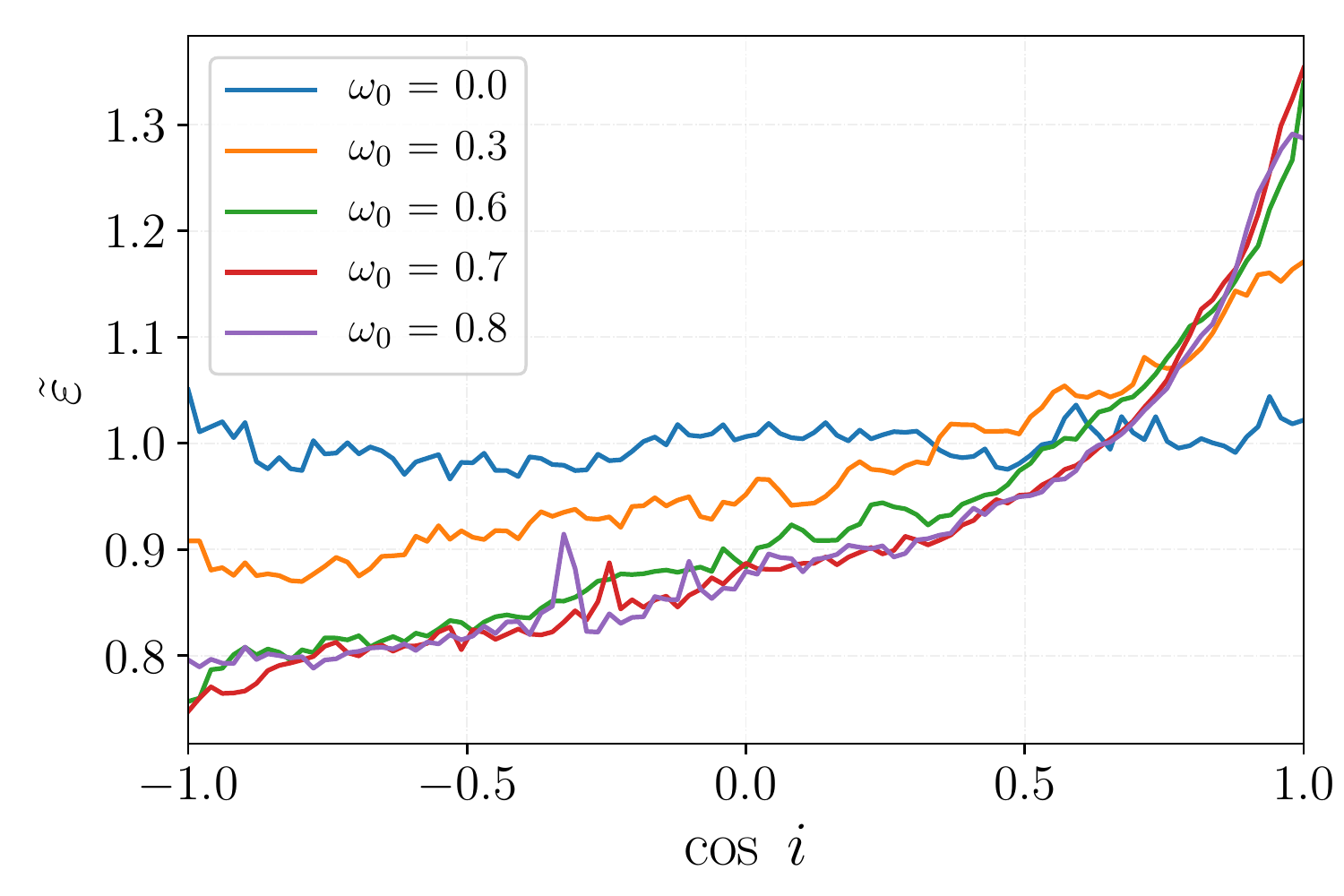}
\caption{\label{fig:3} Radially averaged effective enhancement of average mass as a function of inclination; see Equation (\ref{eq:3}). Curves represent 
the following models: blue is nonrotating $(\omega_0 = 0)$, orange, green, red, and purple are rotating with $\omega_0 = 0.3, 0.6, 0.7$, and $0.8$, respectively. Similarly to Figure~\ref{fig:1}, multiple snapshots are superimposed.}
\end{figure}
We compare the result of the model shown in Figure~\ref{fig:1} with other King models with different rotational parameters by measuring how $\tilde{\varepsilon}(\cos i)$ depends on the rotation of clusters. In Figure~\ref{fig:3}, the averaged curves of $\tilde{\varepsilon}(\cos i)$ are shown (as in the above subsection, the snapshots between 400 and 1000 time units are stacked to improve the statistics) for $N=64\mathrm{k}$ models with $W_0=6$, an $m^{-2}$ mass function, and $\omega_0 = 0,0.3,0.6,0.7,0.8$. The $\tilde{\varepsilon}(\cos i)$ is flat for the nonrotating King model, which implies that there is no anisotropic mass segregation in the (spherically symmetric) nonrotating King models.\footnote{Since in the $N$-body realization $\omega_0=0$ model the net angular momentum is not exactly zero due to Poisson fluctuations. We define the inclination with respect to the plane defined by the residual angular momentum.} Faster rotation leads to a more prominent anisotropic mass segregation. This fact can be seen as the slopes of the $\tilde{\varepsilon}(\cos i)$ curves get steeper for higher values of $\omega_0$. For the highest values of $\omega_0$ shown in this work (between 0.6 and 0.8) the effect saturates, and models with $\omega_0 = 0.9$ and above (not shown in the figure) a bar instability occurs in the simulations \citep{Hong2013}.
\begin{figure}[b!]
\includegraphics[width=1\columnwidth]{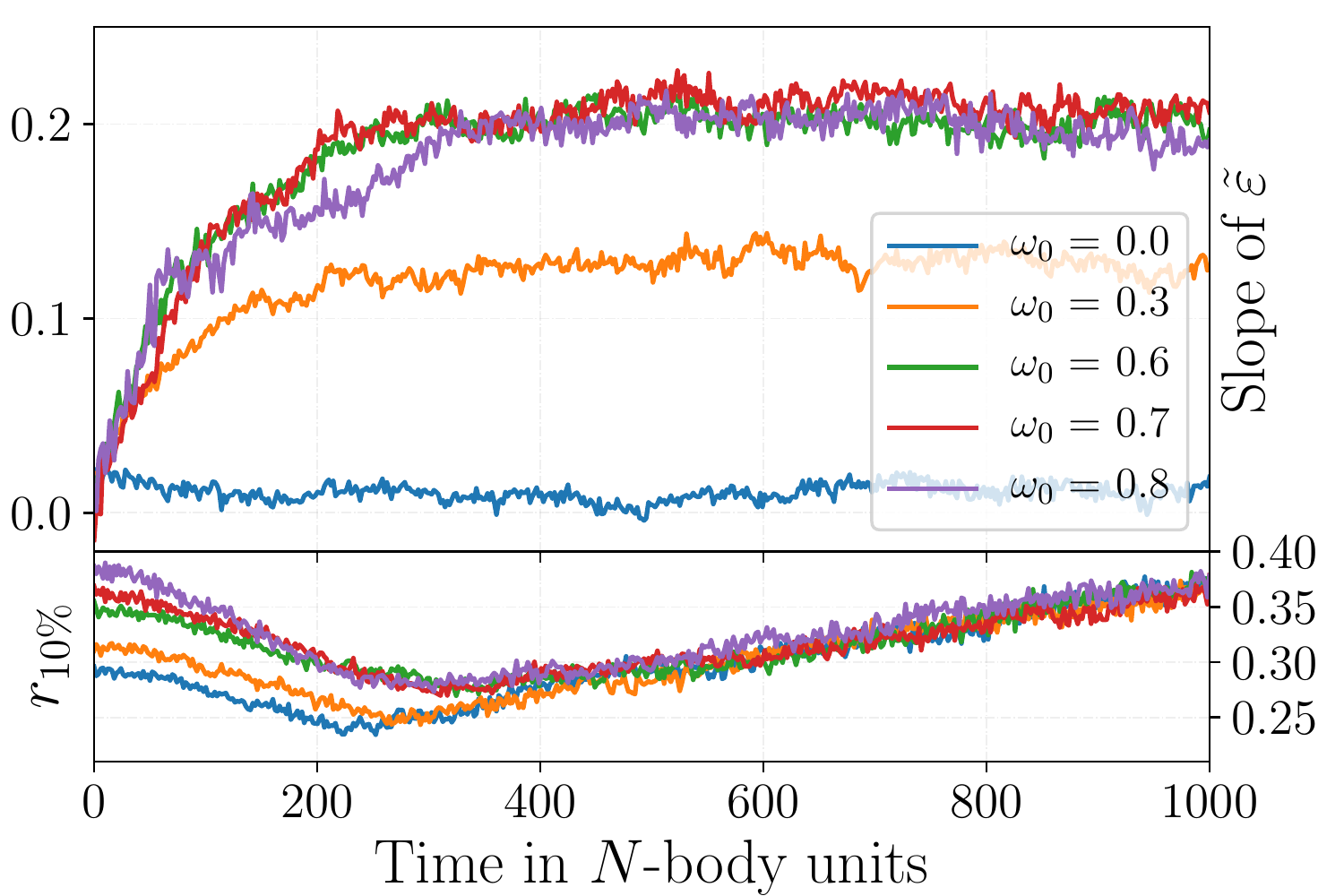}
\caption{\label{fig:4}Best-fitting linear slopes to the effective enhancements $\tilde{\varepsilon}$ as a function of $\cos i$ as a function of time (upper panel). Different colors represent different rotation parameters as in Figure~\ref{fig:3}. The saturation of the curves represents the statistical equilibrium state with respect to anisotropic mass segregation. In the lower panel, the curves are the $10\%$ Lagrangian radii as functions of time for comparison. Anisotropic mass segregation saturates on the core collapse timescale.}
\end{figure}

\newpage
To determine the timescale of anisotropic mass segregation to reach equilibrium from the adopted initial conditions, we measure how the fitted slope of $\tilde{\varepsilon}(\cos i)$ varies with time, as shown in Figure~\ref{fig:4}. When the slope of $\tilde{\varepsilon}(\cos i)$ approaches a constant value, as a function of time, the distribution of inclinations reaches the equilibrium of anisotropic mass segregation. In Figure~\ref{fig:4}, saturation is reached around $300$ H\'enon time units in all simulations with different $\omega_0$ values. We have found that this happens on the core collapse time scale which is indicated by the contraction of the $10\%$ Lagrangian radii as functions of time, see the lower panel on Figure~\ref{fig:4}.

\subsection{The Particle Number Dependence}

\begin{figure}[h!]
\includegraphics[width=1\columnwidth]{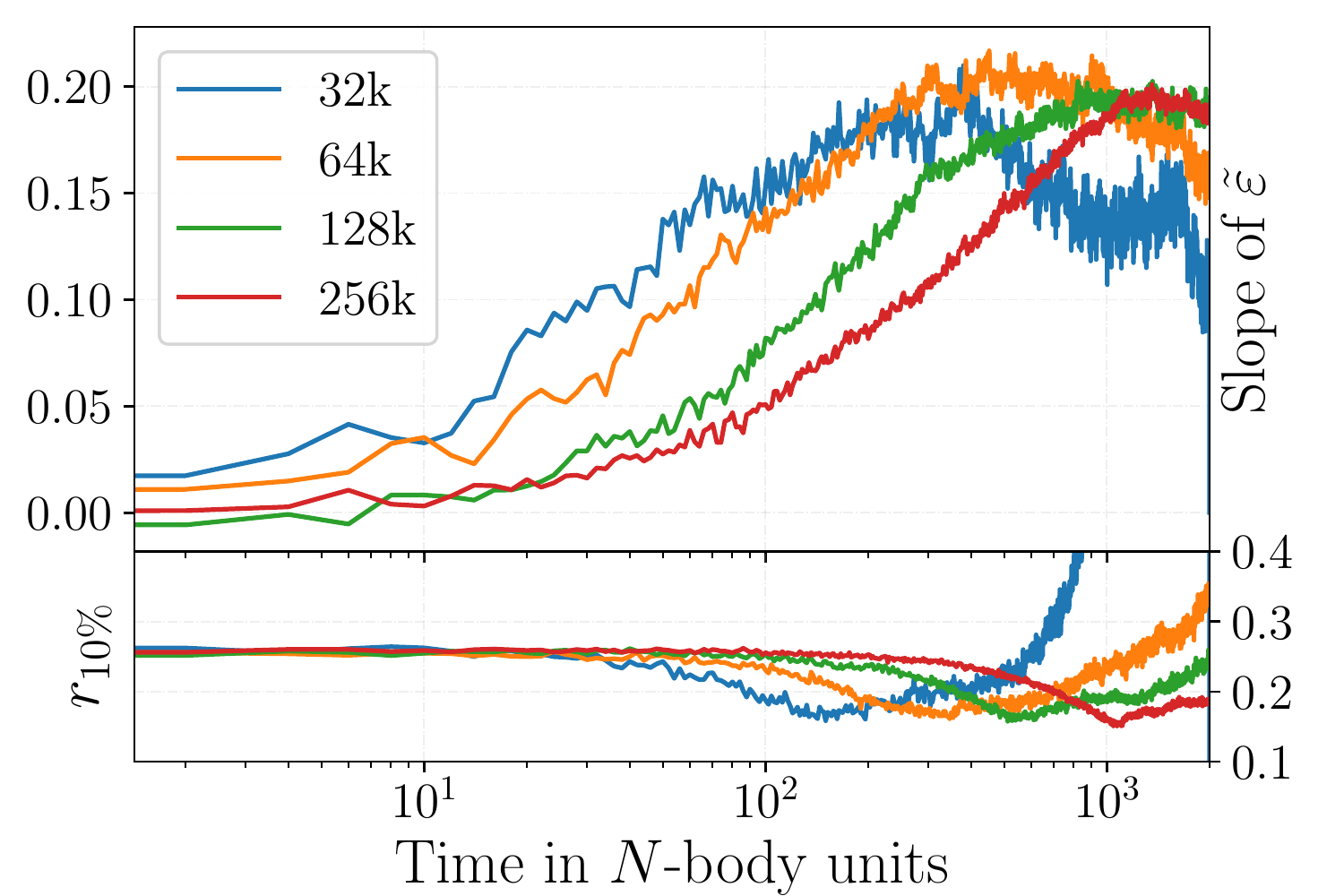}
\caption{\label{fig:5} Same as Figure~\ref{fig:4} but for simulations with 32k (blue), 64k (orange), 128k (green), and 256k (red) stars with $\omega_0 = 0.6$ rotation in each case.}
\end{figure}

We also investigate the efficiency of anisotropic mass segregation with respect to the number of stars in a GC with the same $\omega_0 = 0.6$ rotation. In Figure~\ref{fig:5}, results show the rate of anisotropic mass segregation is roughly linear with $N$, similarly to two-body relaxation.

Longer simulations (with respect to the half-mass relaxation time) show that after reaching the equilibrium state of anisotropic mass segregation, the effect is slowly reduced (i.e., the heavier objects start to isotropize). This is clearly seen for the $N=32$k and $64$k models in Figure~\ref{fig:5}. We found that the saturation timescale of anisotropic mass segregation coincides with the timescale of core collapse.
The final degree of anisotropic mass segregation (at equilibrium) is independent of the number of stars in these simulations at a fixed net rotation, the slope of $\tilde{\varepsilon}$ curves all saturate at around $0.2$ for $\omega_0=0.6$. However, a significant amount of anistropic mass segregation is present already well before core collapse.

\newpage

\subsection{The Effect of the Mass Function}

\begin{figure}[h!]
\includegraphics[width=1\columnwidth]{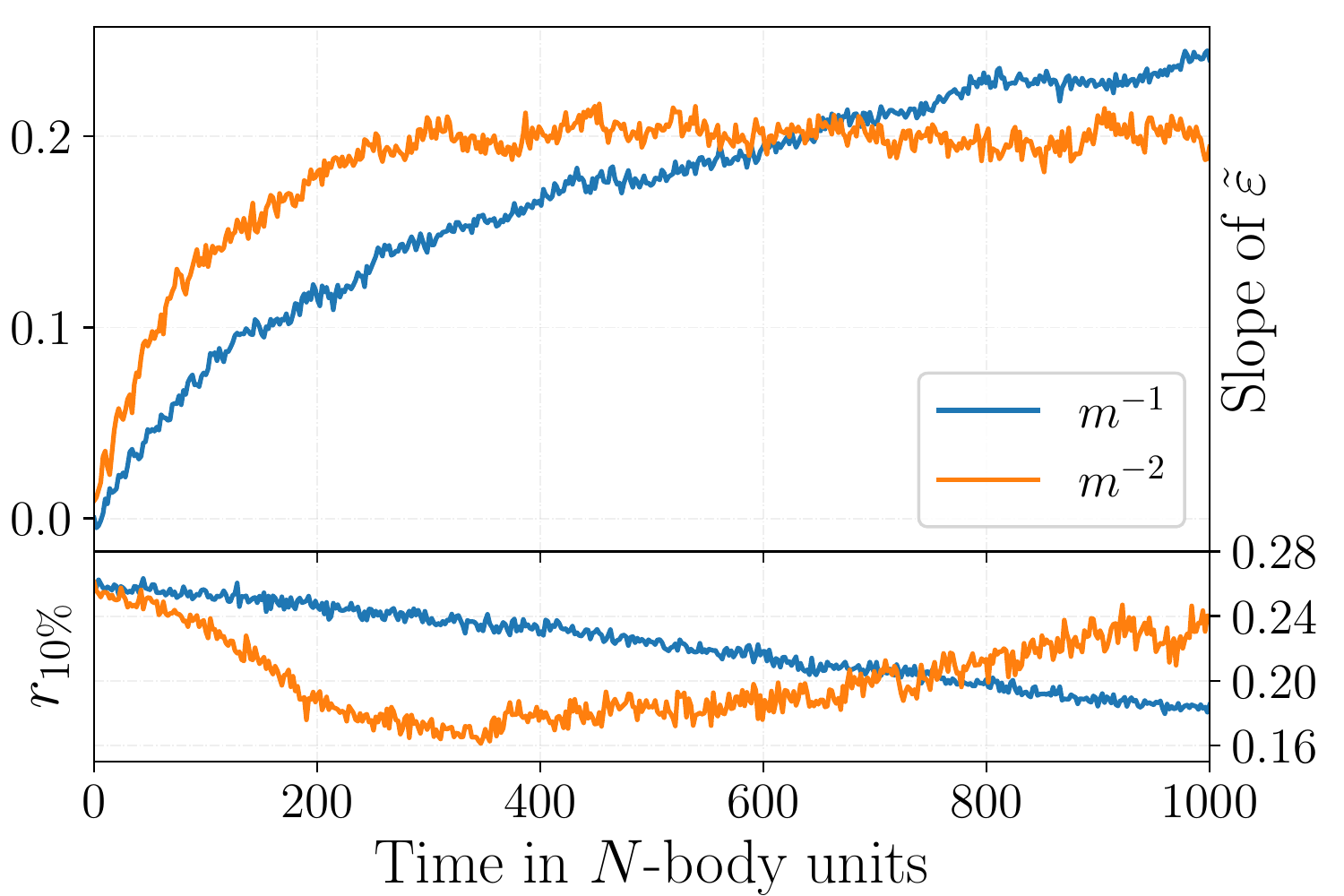}
\caption{\label{fig:6} Similar to} Figure \ref{fig:4} and~\ref{fig:5} but for M64.6.1 and M64.6.2 models with $m^{-1}$ (blue) and $m^{-2}$ (orange). While the latter model reaches the equilibrium at $t\sim300$, the former model does not within the duration of the simulation (1000 time units).
\end{figure}

We compare rotating King models with stellar mass distribution $p(m) \propto m^{-1}$ (bottom-heavy model) and $p(m) \propto m^{-2}$ (top-heavy model) to examine the dependence of anisotropic mass segregation on the mass function. These models have the same rotation parameter of $\omega_0=0.6$ and particle number $N=64\mathrm{k}$. The ratio between the sum-squared mass of the bottom-heavy and top-heavy model is $\sim 0.53$. Figure~\ref{fig:6} shows that the saturation value, where anisotropic mass 
segregation reaches its equilibrium, correlates with the number of heavy objects in a cluster. While the redistribution of inclinations approaches the equilibrium within $t\sim300$ time units for the top-heavy model, the bottom-heavy model has not saturated yet at $t = 1000$. Figure~\ref{fig:6} also shows that anisotropic mass segregation appears on the same timescale as core collapse for $N=64$k for different mass spectra.

\subsection{Superimposed Counterrotating Clusters}

Finally, we examine what happens when two rotating King models are superimposed such that their total angular momenta are pointing in opposite directions, creating a cluster with no net rotation but an aspherical shape and bimodal distribution of angular momenta. To generate the initial condition, we superimpose two $(W_0, \omega_0)=(6, 0.6)$ models with $N=32\mathrm{k}$ each. 

\begin{figure}[t!]
\includegraphics[width=1\columnwidth]{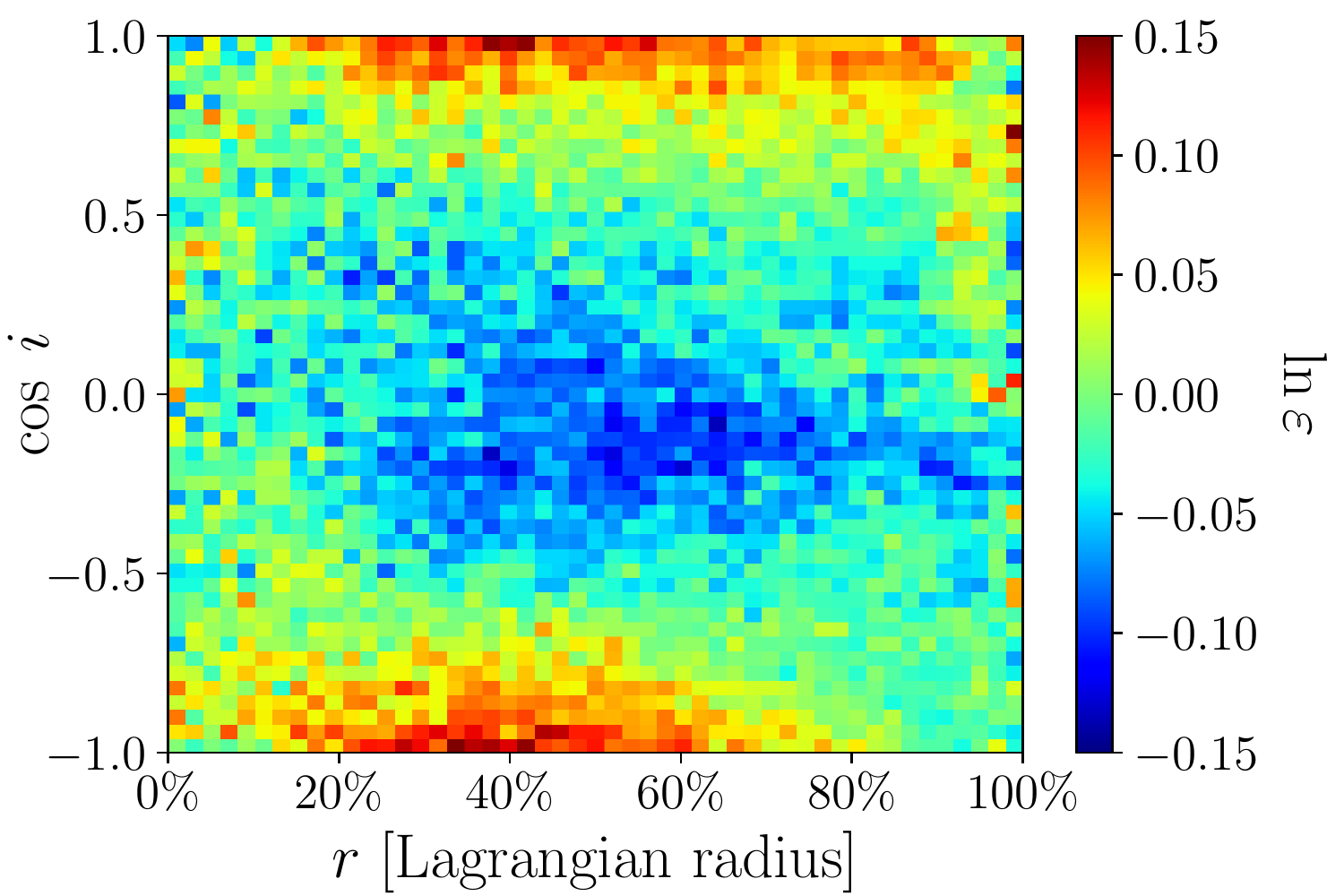}
\caption{\label{fig:7} Same as Figure~\ref{fig:1} but for M64.6.2x model where two subclusters are superimposed such that their total angular momenta are pointing in opposite direction, creating a cluster with no net rotation but an aspherical shape. This cluster is in equilibrium state with respect to the distribution of inclinations.}
\end{figure}

Figure~\ref{fig:7} shows that two counterrotating structures form. The one that has positive rotation is radially more extended, but flattened ($\cos i \gtrsim 0.8$); in contrast, the other one is less extended radially, but has wider spread in inclination ($\cos i \lesssim 0.75$). We note that this can be a result of a statistical fluctuation because the system was initially symmetric with respect to its equatorial plane.

\section{Conclusions} \label{sec:conclusions}

In this paper, we examined the inclination distribution of stellar orbits in direct $N$-body simulations of rotating GCs with mass spectra. We found evidence of an equilibrium state with anisotropic mass segregation. Our results show that there is a statistical enhancement of average mass for the lowest inclination orbits, beyond the well-known radial mass segregation of the heaviest objects. The rotating systems rapidly reach the equilibrium of anisotropic mass segregation, while the clusters are still far from energy equipartition. The clusters did not reach energy equipartition even after several two-body relaxation times in agreement with \citet{Trenti2013}. The average mass can be up to $2.7$ times larger at the lowest inclination orbits ($i\lesssim25^{\circ}$) between the $5$\% and $40$\% Lagrangian radii where the effect is the most prominent; see Figure \ref{fig:1}. This result implies that the distribution of stellar mass black holes represents a thick disk near the centers of rotating GCs. This prediction may have important implications for modeling black hole populations, black hole binary formations, and gravitational wave emission rates in GCs which traditionally assumed to be isotropic (e.g.~\cite{Rodriguez2016,Askar2016,Wang2016,Park2017}). If black holes follow a flattened distribution in the centers of GCs, this increases their number density and decreases their velocity dispersion, which may imply a higher dynamical encounter rate, and lead to an enhanced black hole binary formation rate. Calculating the binary formation rate, as well as predicting the gravitational merger rate, are beyond the scope of the current work.

We emphasize that while anisotropic mass segregation reaches a steady state, the cluster itself actually does not. The efficiency of anisotropic mass segregation at equilibrium depends on the degree of rotation, and the mass distribution in the cluster. The time it takes for anisotropic mass segregation to reach equilibrium in $N$-body units is approximately linear in the number of stars and also depends on the mass distribution. We found that this timescale coincides with the core collapse timescale in simulations, although, a significant amount of anistropic mass segregation is present already well before core collapse (Figure~\ref{fig:5}).
Anisotropic mass segregation affects the distribution of high mass objects most prominently (Figure~\ref{fig:2}), and takes place well within the half-mass radius (i.e. between $0$\% and $60$\% Lagrangian radii; see Figure~\ref{fig:1}), where two-body relaxation is more efficient than in the outer regions; see Figure 3 in \citep{Meiron2018}. In future work, we plan to explore how stellar evolution, binary evolution, galactic tidal forces and possible mergers affect the appearance and evolution of anisotropic mass segregation in rotating stellar clusters. Further analysis is also needed to determine the relative contribution of two-body relaxation and VRR in driving anisotropic mass segregation in GCs.

\acknowledgments{
We thank Scott Tremaine, Rainer Spurzem, Peter Berczik, Manuel Arca Sedda and Jongsuk Hong for helpful comments. This work received funding from the European Research Council (ERC) under the European Union's Horizon 2020 research and innovation programme under grant agreement No.~638435 (GalNUC) and was supported by the Hungarian National Research, Development, and Innovation Office grant NKFIH KH-125675. Yohai Meiron acknowledges support from an NSERC grant to Ray Carlberg. The calculations were carried out on the NIIF HPC cluster at the University of Debrecen, Hungary.}

\bibliography{szolgyen_et_al}

\end{document}